\documentclass[aps,prb,reprint,superscriptaddress,showpacs]{revtex4-1}

\usepackage{amsmath,amssymb,amsfonts}
\usepackage{graphicx}
\usepackage{hyperref}
\usepackage{braket}
\usepackage{bm}

\usepackage{comment}

\begin{document}
\title{Renormalization group study of electromagnetic interaction in multi-Dirac-node systems}
\author{Hiroki Isobe}
\affiliation{Department of Applied Physics, University of Tokyo, Tokyo 113-8656, Japan}
\author{Naoto Nagaosa}
\affiliation{Department of Applied Physics, University of Tokyo, Tokyo 113-8656, Japan}
\affiliation{RIKEN Center for Emergent Matter Science, ASI, RIKEN, Wako, Saitama 351-0198, Japan}
\date{\today}

\begin{abstract}
We theoretically study the electromagnetic interaction in 
Dirac systems with $N$ nodes by using the 
renormalization group, which is relevant to the quantum critical phenomena 
of topological phase transition ($N=1$) and Weyl semimetals ($N=4$ or $N=12$). 
Compared with the previous work for $N=1$ [H. Isobe and N. Nagaosa, Phys. Rev. B {\bf 86}, 165127 (2012)],
we obtained the analytic solution for the large $N$ limit, which differs qualitatively 
for the scaling of the speed of light $c$ and that of electron $v$, i.e., $v$ does not
change while $c$ is reduced to $v$.   
We also found a reasonably accurate approximate analytic solution for 
generic $N$, which well interpolates between $N=1$ and large $N$ limit, 
and it concludes that $c^2 v^N$ is almost unrenormalized.
The temperature dependence of the physical properties,  
the dielectric constant,  magnetic susceptibility, spectral function, 
DC conductivity, and mass gap are discussed based on these results. 
\end{abstract}
\maketitle

\section{Introduction}

Dirac fermions are spin 1/2 particles described by the basic equation of the relativistic quantum mechanics, 
the Dirac equation.  
\cite{*[{For a description of the Dirac equation and QED, see for example }] 
peskin1995itq, *ramond1990ftm}
Since it is based on the special relativity, the Dirac equation is invariant under the Lorentz 
transformation. Dirac fermions are described by four-component spinors, and their components correspond 
to positive and negative energy and spin freedom. When the mass of a Dirac fermion is nonzero, the 
four-component representation is irreducible, but in the massless case, it becomes reducible to be 
a two-component representation. This two-component fermion is called a Weyl fermion. There exists the chiral 
symmetry for Weyl fermions, so they can be distinguished by the chirality. Right-handed or left-handed Weyl 
fermions cannot exist independently, thus the number of Weyl fermions is always even. This is the result of the 
fermion doubling theorem. \cite{nielsen1983abj}

The interaction between Dirac fermions and electromagnetic field is formulated in quantum electrodynamics 
(QED), and the exchange of photons mediates the interaction force. In QED, the speed of electron $v$ 
and that of light $c$ has the same value, and QED is the Lorentz-invariant theory. QED is also known as the most 
precise theory in physics. 

The electronic states in solids are described by the Bloch wave functions, and according to the band theory, 
the equation equivalent to the Dirac equation may appear. One such example is graphene, a two-dimensional 
carbon sheet forming hexagonal lattice. \cite{neto2009tep} 
The effective theory is described by the $2 \times 2$ Dirac Hamiltonian, 
and Dirac spectra appear at $K$ and $K'$ points in the Brillouin zone. Another example is bismuth, 
which exhibits a four-component massive Dirac fermion caused by spin-orbit interaction. 
\cite{[{}] [{ and references therein.}] fuseya2009icf}
Topological insulators also have Dirac spectrum on the surface. 
\cite{hasan2010cti,qi2011tia} 
Although the bulk is 
insulating and gapped in topological insulators, the gap closes at the quantum phase transition between 
topological and trivial insulators. The effective theory at the critical point is described by the $4 \times 4$ 
Dirac Hamiltonian for the systems with inversion symmetry, 
and the sign change of mass $m$ corresponds to the phase transition. Namely, in this case
the number of the Dirac fermion $N$ is 1.
This scenario is experimentally confirmed in BiTl(S$_{1-x}$Se$_x$)$_2$ by changing the concentration $x$. 
\cite{xu2011tpt,sato2011uma} 
Pyrochlore iridates are predicted to be Weyl semimetals, where Weyl nodes are located on the Fermi surface. 
Band calculations indicate that there are $2N=24$ (or $2N=8$) Weyl nodes exist in pyrochlore iridates. 
\cite{wan2011tsa,witczak-krempa2012tam}

When a Dirac point is located on the Fermi level, the electron-electron interaction is not well screened, 
and becomes a long-range force. Thus, the effective model for Dirac fermion in solids has nearly equivalent 
form to QED. The renormalization group (RG) is used to deal with divergent integrals appearing in the 
perturbative treatment of the interaction. Here we should note one important difference from QED. 
In the band theory, the group velocity of electron $v$ is expressed by the derivative of the energy dispersion 
in terms of the crystal momentum, but this $v$ is far smaller than the speed of light in solid $c$. 
Therefore, the Lorentz invariance is broken in this model. The smallness of the factor $v/c$ naturally leads 
to the choice of Coulomb gauge, where the scalar potential gives the instantaneous Coulomb interaction 
though the transverse part of the vector potential is often neglected. 

The effects of electron-electron interaction on Dirac electrons are extensively studied. 
\cite{kotov2012eei,goswami2011qcba,hosur2012cti,gonzalez1994nfl}
RG analyses of Dirac electrons considering instantaneous Coulomb interaction in two and three dimensions 
\cite{kotov2012eei,goswami2011qcba,hosur2012cti} 
reveal the logarithmic divergence of $v$, while the coupling constant $\alpha$ is marginally irrelevant. 
In these analyses, $c$ is not renormalized and stays constant. 
The divergence of $v$ contradicts the assumption that the factor $v/c$ is small. 
When the contribution from the vector potential is considered for 2D system, 
$v$ saturates to $c$, and the Lorentz invariance is recovered in the low-energy limit. \cite{gonzalez1994nfl}
The quantum critical behavior of Dirac electrons close to the superconducting transition is also studied. 
\cite{roy2013qsc}
For 3D system, the previous study for $N=1$~\cite{isobe2012tqca} reveals the renormalization of $c$ in addition to $v$, 
and the recovery of Lorentz invariance in the low-energy limit.  
However, as mentioned above, there are cases where the number of Dirac nodes $N$ 
takes several values in condensed matter systems, and it is important to extend the analysis to 
the generic $N$. 

In this paper, we study the electromagnetic interaction in multinode Dirac and Weyl systems. 
Especially, in the limit of large $N$,  we can obtain the analytic solution to RG
equations. For generic $N$, we found a reasonably accurate approximate 
solution that interpolates the two limits, i.e., $N=1$ and large $N$. By these results,
we have revealed the global view of the RG flow in this problem and
clarified the condition for neglecting the transverse channels of the electromagnetic 
interaction. We also made corrections of the previous study. \cite {isobe2012tqca}

\section{Model}
In this study we use the following Lagrangian: 
\begin{align}
\label{eq:lagrangian}
\mathcal{L}= \ 
	& \bar{\psi}_a (\gamma^0 p_0-v\bm{\gamma}\cdot\bm{p}-m)\psi_a
	+\frac{1}{2}(\varepsilon \bm{E}^2-\frac{1}{\mu}\bm{B}^2) \notag \\
	&-e\bar{\psi}_a l^\mu_\nu\gamma^\nu \psi_a A_\mu.
\end{align}
Here $\psi_a$ is a four-component Dirac spinor with $a$ being the $N$ node index. 
The matrix 
\begin{equation}
l^\mu_\nu =
\begin{pmatrix}
1 & & & \\
& v/c & & \\
& & v/c & \\
& & & v/c 
\end{pmatrix}
\end{equation} 
is introduced to describe the electromagnetic interaction in a system without Lorentz invariance. 
The metric used in the model is $(+ - - -)$. 
We mostly focus on the massless case $(m=0)$, which corresponds to a quantum critical point of 
topological insulators and Weyl semimetals. 
The RG effect on $m$ is discussed later. 

To accurately describe the topological insulator phase, the $\theta$ term is necessary; 
i.e., we should add $\theta \bm{E} \cdot \bm{B}$ $(\theta = \pm \pi)$ to the action. 
The $\theta$ term is omitted in the model because this term can be transformed into the surface term 
and we do not consider the topological magneto-electric effect in the present analysis. 
Actually, the RG analysis does not modify the $\theta$ term.
It is natural since topological terms have discrete integer values, and we confirmed 
this fact from the following two methods: 
the perturbative calculation and the background field theory. 
In any case, the topological $\theta$ term does not alter the bulk properties.  

If we consider massless Weyl nodes, the Lagrangian has the chiral symmetry, and 
a four-component Dirac spinor can be separated into two two-component 
Weyl spinors with opposite chiralities. 
Thus, the number of Weyl nodes $N_\text{W}$ are twice as large as that of 
Dirac nodes $N$, i.e., $N_\text{W} = 2N$.
In the following analysis, we treat the model in the four-component notation. 
If necessary, we can use the projection operator $(1\pm\gamma^5)/2$ to separate 
a massless Dirac fermion into two Weyl fermions with opposite chiralities. 

The permittivity $\varepsilon$ and the permeability $\mu$ determine the speed of light in material 
$c=c_\text{vacuum}/\sqrt{\varepsilon \mu}$, where $c_\text{vacuum} = 3 \times 10^8\, \mathrm{m/s}$ 
is that in vacuum. $c$ also sets the speed of electromagnetic interaction in material. 
The electric and magnetic fields are written as 
\begin{equation}
\label{eq:def_emf}
\bm{E}=-\frac{1}{c}\frac{\partial\bm{A}}{\partial t}
-\bm{\nabla} A_0, 
\ \ 
\bm{B}=\frac{1}{c}\bm{\nabla} \times \bm{A}.
\end{equation}
The electron propagator $G_0(p)$, the photon propagator $D^{\mu\nu}_0(p)$, 
and the vertex $\Gamma_0^\mu$ are given by
\begin{gather}
G_0(p) = \frac{i}{\gamma^0 p_0 +v \gamma^\alpha p_\alpha +i0}, \\
D^{\mu\nu}_0(q) = \frac{-ic^2g^{\mu\nu}}{\varepsilon(q_0^2-c^2q_\alpha^2)+i0}, \\
\Gamma_0^\mu = -ie l^\mu_\nu \gamma^\nu = 
	\left( -ie\gamma^0 , -ie\frac{v}{c}\gamma^\alpha \right) .
\end{gather}
We have used the Feynman gauge for the photon propagator. Physical quantities are independent of 
the gauge choice.

\section{Renormalization group analysis}

\begin{figure}
\centering
\includegraphics[width=0.9\hsize]{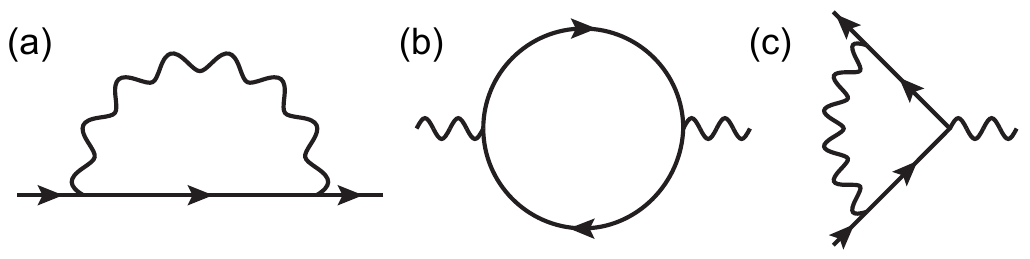}
\caption{Feynman diagrams at one-loop order: (a) self-energy, (b) polarization, (c) vertex.}
\label{fig:diagram3}
\end{figure}

We calculated the Feynman diagrams to one-loop order (Fig.~\ref{fig:diagram3}) to derive the RG equations: 
\begin{gather}
\kappa\frac{dv}{d\kappa} = -\frac{2g^2}{3\pi}\frac{c^2}{(c+v)^2}
	\left[ 1+2\left( \frac{v}{c} \right)+\left( \frac{v}{c} \right)^2 
	-4\left( \frac{v}{c} \right)^3 \right] , \label{eq:rg_v} \\
\kappa\frac{dc}{d\kappa} = \frac{Ng^2}{3\pi}
	\frac{c}{v} \left[ 1- \left( \frac{v}{c} \right)^2 \right] , \label{eq:rg_c} \\
\kappa\frac{dg^2}{d\kappa} =  \frac{2Ng^4}{3\pi}\frac{1}{v}. \label{eq:rg_g2} 
\end{gather}
Here, $\kappa$ denotes the momentum scale. 
The details of the calculation for $N=1$ are described in the previous paper. \cite {isobe2012tqca}

\subsection{Large $N$ limit}

We set a parameter $\lambda$, which measures the effect of a polarization bubble compared to the bare propagator, as 
\begin{equation}
\lambda = N\alpha = \frac{Ng^2}{v}.
\end{equation}
If we write the RG equations in terms of $\lambda$ instead of $g$, we obtain 
\begin{align}
\frac{dv}{dl} &= \frac{2\lambda}{3\pi N} \frac{c^2 v}{(c+v)^2} 
	\left[ 1 + 2 \left( \frac{v}{c} \right) + \left( \frac{v}{c} \right)^2 
	-4 \left( \frac{v}{c} \right)^3 \right], \\
\frac{dc}{dl} &= -\frac{\lambda}{3\pi} c \left[ 1- \left( \frac{v}{c} \right)^2 \right], \\
\frac{d\lambda}{dl} &= -\frac{2\lambda^2}{3\pi} - \frac{2\lambda^2}{3\pi N} 
	\left[ 1 + 2 \left( \frac{v}{c} \right) + \left( \frac{v}{c} \right)^2 
	-4 \left( \frac{v}{c} \right)^3 \right].
\end{align}
where we define $l=\ln(\kappa_0/\kappa)$. 
Terms proportional to $1/N$ vanish in the large $N$ limit ($N \to \infty$), and the analytical solutions to 
the RG equations are easily obtained as  
\begin{align}
v(l) &= v_0 \ \text{(const.)}, \\
c(l) &= \left[ v_0^2 + (c_0^2 - v_0^2) \left( 1 + \frac{2\lambda_0}{3\pi} l \right)^{-1} \right]^{1/2}, 
	\label{eq:c_largeN} \\
\lambda (l) &= \lambda_0 \left( 1 + \frac{2\lambda_0}{3\pi} l \right)^{-1}. 
\end{align}
From Eq.~\eqref{eq:c_largeN}, we can confirm 
$c(l) \to v_0$ $(l \to \infty)$, while $v$ is not renormalized.  
This is in sharp contrast to the result obtained by neglecting the transverse electromagnetic field, 
where $v$ diverges logarithmically.  
It means the recovery of the Lorentz invariance in the infrared (IR) limit.

\subsection{Numerical solutions}
\begin{figure}
\centering
\includegraphics[width=0.9\hsize]{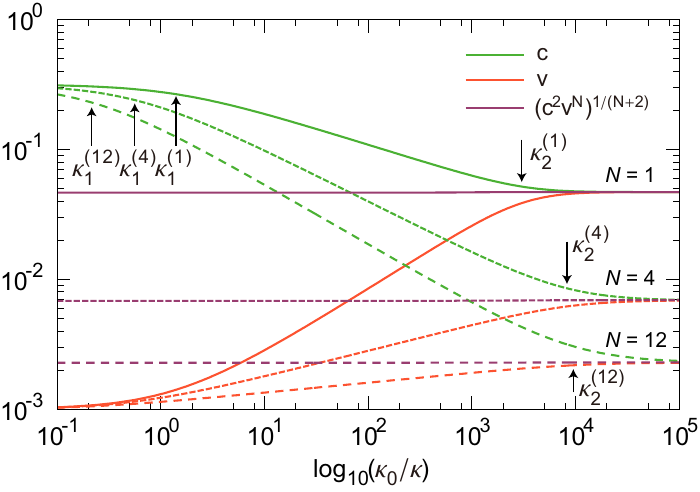}
\caption{%
(Color online) 
Numerical solutions to the RG equations for $v$ and $c$. 
We set the initial values $v_0=0.001$ and $\varepsilon_0=10$. 
A nonmagnetic material ($\mu_0 =1$) is considered, and in this case, 
$c_0=0.32$ and $\alpha_0=0.73$. 
We can observe that the quantities $(c^2v^N)^{1/(N+2)}$ are almost constant 
for all momentum scale. 
}
\label{fig:rg1}
\end{figure}

\begin{figure}
\centering
\includegraphics[width=0.9\hsize]{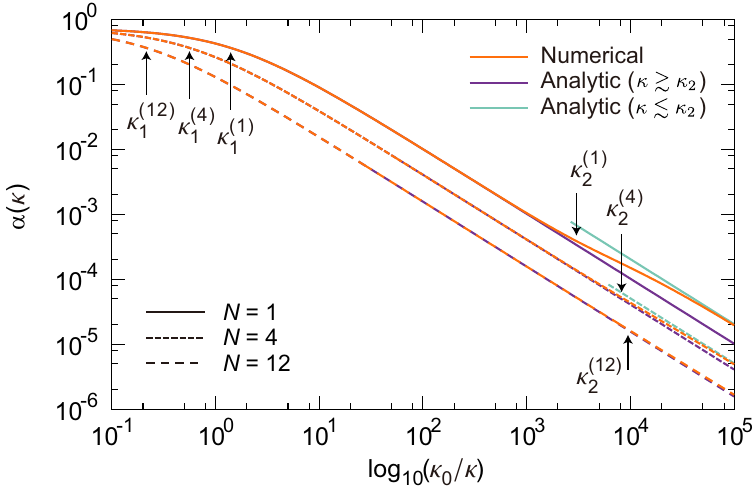}
\caption{%
(Color online) 
Numerical and analytic solutions to the RG equations for $\alpha$. 
The analytic solutions for $\kappa\gtrsim\kappa_2$ and $\kappa\lesssim\kappa_2$ 
well explain the numerical solutions.
As the number of species $N$ increases, the difference between the analytic solutions 
for $\kappa\gtrsim\kappa_2$
and numerical solutions becomes smaller, and the analytic expression becomes more precise. 
}
\label{fig:rg_alpha}
\end{figure}

The RG equations~\eqref{eq:rg_v}, \eqref{eq:rg_c}, and \eqref{eq:rg_g2} for generic $N$ 
cannot be solved analytically without any approximations, so we first solve them numerically.  
The numerical solutions to the RG equations are shown 
in Figs.~\ref{fig:rg1} and \ref{fig:rg_alpha}. 
We set the initial (bare) values of $v_0=0.001$ and $\varepsilon_0 = 10$ and 
consider a nonmagnetic material ($\mu_0=1$). 
In this case, $c_0=0.32$ and $\alpha_0=0.73$, where 
the dimensionless coupling constant $\alpha$ is defined by 
\begin{equation}
\alpha=\frac{g^2}{v}=\frac{e^2}{(4\pi\varepsilon) v}.
\end{equation}

We can find some important features from the result. 
First, the speed of electron $v$ and that of light $c$ coincide to be the common value 
$c_\infty = (c_0^2 v_0^N)^{1/(N+2)}$ in the IR limit. 
Second, the quantity $c^2 v^N$ is almost constant for all momentum scale. 
We make use of this fact for the analytical solutions discussed below. 
Third, the dimensionless coupling constant $\alpha$ becomes smaller in the IR limit, which 
concludes the validity of the perturbative RG analysis. 
Therefore, the Lorentz invariance is recovered in the IR limit, and the system becomes 
equivalent to that of the conventional QED. 
Even if the Lorentz invariance is broken in the original Lagrangian, the RG analysis reveals 
that the system in the IR limit is the ideal laboratory to study QED.

\subsection{Analytic solutions}

As we saw in the numerical calculations, the quantity $c^2 v^N$ is almost constant 
independent of the momentum scale. 
From the RG equations, the scale dependence of the quantity is 
\begin{align}
\label{eq:approx11}
\frac{d(c^2v^N)}{dl} =& 2cv^N\frac{dc}{dl} + Nc^2v^{N-1}\frac{dv}{dl} \notag \\
	=& \frac{2Ng^2}{3\pi} c^2v^{N-1} \frac{\beta^2(1-\beta)^2}{(1+\beta)^2},
\end{align}
where $\beta=v/c$. 
If we define the function $f(\beta)$ as 
\begin{gather}
f(\beta) = \frac{\beta^2(1-\beta)^2}{(1+\beta)^2}, 
\end{gather}
and assume $v \leq c$, i.e. $0 \leq \beta \leq 1$, we obtain 
$0 \leq f(\beta) \leq 17-12\sqrt{2} \simeq 0.03$.
The maximum value $g(\beta) \simeq 0.03$ is rarely observed 
in the scale of Fig.~\ref{fig:rg1}, and
the right-hand side of Eq.~\eqref{eq:approx11} is always small for 
$0<\beta<1$. Therefore, the approximation 
\begin{equation}
\label{eq:approx1}
c^2v^N = c_0^2v_0^N
\end{equation}
is satisfied for the entire energy scale.

The second approximation is 
\begin{equation}
\label{eq:approx2}
\frac{c}{c_0} = \frac{g}{g_0}. 
\end{equation}
It holds until $c$ reaches the vicinity of the asymptotic value $c_\infty$.  
Actually, this approximation has a physical interpretation. 
Since $c=1/\sqrt{\varepsilon\mu}$ and $g=e/\sqrt{4\pi\varepsilon}$, 
the equality means the permeability $\mu$ stays constant. 

Using Eqs.~\eqref{eq:approx1},~\eqref{eq:approx2}, we can analytically solve the 
RG equations~\eqref{eq:rg_v}, \eqref{eq:rg_c}, and \eqref{eq:rg_g2}, and obtain 
\begin{equation}
\label{eq:analy_g}
g^2(l) = g^2_0 \left( 1+ \frac{2N+2}{3\pi}\alpha_0 l \right)^{-N/(N+1)}.
\end{equation}
The other solutions follow by using the analytic expression of $g^2(l)$ as  
\begin{gather}
v(l) = v_0 \left( 1+ \frac{2N+2}{3\pi}\alpha_0 l \right)^{1/(N+1)}, \\
c(l) = c_0 \left( 1+ \frac{2N+2}{3\pi}\alpha_0 l \right)^{-N/(2N+2)}, \\
\alpha(l) = \alpha_0 \left( 1+ \frac{2N+2}{3\pi}\alpha_0 l \right)^{-1}.
\end{gather}
These analytic expressions are valid for $\kappa \gtrsim \kappa_2$. 

From the analytical solutions, 
we can identify the two momentum scales, $\kappa_1$ and $\kappa_2$, as
\begin{subequations}
\begin{gather}
\kappa^{(N)}_1 = \exp\left[ -\frac{3\pi}{(2N+2)\alpha_0} \right] \kappa_0, \\
\kappa^{(N)}_2 = \exp\left[ -\frac{3\pi}{(2N+2)\alpha_0} 
	\bigg[ \left( \frac{c_0}{v_0} \right)^{(2N+2)/(N+2)} -1 \bigg] \right] \kappa_0.
\end{gather}
\end{subequations}
$\kappa_1$ is determined by $\alpha(\kappa_1) = \alpha_0/2$ and 
$\kappa_2$ is the point 
where the analytically derived function $c(\kappa)$ coincides with the asymptotic value 
$c_\infty$. 
Assuming $v_0/c_0 \ll 1$, $\kappa_2 \ll \kappa_1 < \kappa_0$ is satisfied. 
These characteristic momenta specify three scaling regions: 
(i) perturbative region $\kappa_1\lesssim \kappa \lesssim \kappa_0$, 
where the deviation from the bare value is small and it can be treated perturbatively; 
(ii) nonrelativistic scaling region $\kappa_2 \lesssim \kappa \lesssim \kappa_1$,  
where the effect of RG becomes large, while the factor $v/c$ is still small; and 
(iii) relativistic scaling region $\kappa \lesssim \kappa_2$, 
where $c(\kappa) \simeq v(\kappa)$ and the Lorentz invariance is recovered. 

As to the dimensionless coupling constant $\alpha$, its analytic expression can be 
obtained for region (iii), the relativistic scaling region. 
When we put $c=v=c_\infty$, the RG equation for $\alpha$ becomes 
\begin{equation}
\frac{d\alpha}{dl} = -\frac{2N}{3\pi}\alpha^2, 
\end{equation}
and it can be solved analytically to obtain 
\begin{equation}
\label{eq:alpha3}
\alpha(l) = \frac{3\pi}{2N} \frac{1}{l}. 
\end{equation}
Surprisingly, the coupling constant $\alpha(l)$ in region (iii) is independent of 
its bare value $\alpha_0$. 

\section{Physical Properties}
\subsection{Density of states}

\begin{figure}
\centering
\includegraphics[width=0.9\hsize]{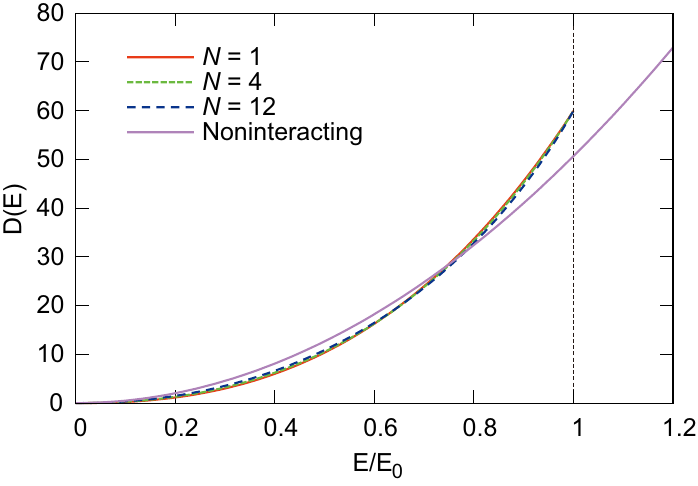}
\caption{%
(Color online) 
Density of states modified by the RG analysis. 
The DOS in the low-energy region is suppressed compared to the noninteracting one, 
due to the electron correlation effect. 
To compensate for the suppression, the DOS increases around $E/E_0 \gtrsim 0.8$. 
The effect of RG appears only below the cutoff energy $E_0$. 
}
\label{fig:RG_DOS}
\end{figure}

The density of states (DOS) is an important quantity to determine the physical property of a material. 
From the RG analysis, the electron velocity $v(k)$ is not a constant, and 
the energy $E(k) = v(k)k$ is no longer linear in the momentum $k$ below the cutoff. 
In general, the DOS of a system with energy $E(k)$ is determined as 
\begin{equation}
D(E) = \int \frac{d^3k}{(2\pi)^3} \delta (E-E(k))
	= \frac{1}{2\pi^2} \frac{k^2(E)}{E'(k(E))}, 
\end{equation}
where $E'$ stands for $dE/dk$. The DOS is a function of energy, so all quantities 
should be expressed in terms of energy $E$. 

The DOS for 3D noninteracting Dirac fermions is 
\begin{equation}
D_0(E) = \frac{E^2}{2\pi^2 v_0^3}.
\end{equation}
The RG effect on the DOS is calculated numerically and is compared with the 
noninteracting case in Fig.~\ref{fig:RG_DOS}. 
Since $v(k)$ gets faster as the momentum scale goes to the IR region, 
the DOS is suppressed in the low-energy region. 
On the other hand, the DOS is increased for $0.8 \lesssim E/E_0 <1$, where 
$E_0 = v_0 \kappa_0$ is the energy cutoff. 
This increase compensates for the suppression of the DOS in the low-energy region.

\subsection{Electromagnetic properties}

\begin{figure} 
\centering
\includegraphics[width=0.9\hsize]{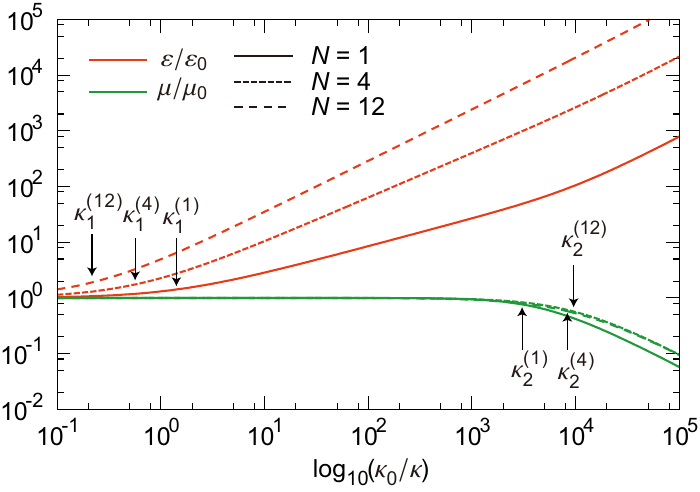}
\caption{%
(Color online) 
Numerical solutions to the RG equations for the permittivity $\varepsilon$ 
and the permeability $\mu$. The characteristic momentum scales for $\varepsilon$ and $\mu$ are 
$\kappa_1$ and $\kappa_2$, respectively.}
\label{fig:emf}
\end{figure}

Here we consider the permittivity $\varepsilon(\kappa)$ and the permeability $\mu(\kappa)$. 
The scale dependence of the permittivity $\varepsilon(\kappa)$ is determined from 
that of $g^2$. We consider that the scale dependence of $g^2$ emerges only from 
$\varepsilon$ and that the bare electric charge $e$ stays constant. 
The permeability is obtained by $\mu=1/(\varepsilon c^2)$. 
The numerical solutions for $\varepsilon$ and $\mu$ are shown in Fig.~\ref{fig:emf}. 
For $\kappa\gtrsim\kappa_2$, the analytic solution to $\varepsilon$ 
is easily obtained from Eq.~\eqref{eq:analy_g} as
\begin{equation}
\varepsilon(l) = 
\varepsilon_0 \left( 1+ \frac{2N+2}{3\pi}\alpha_0 l \right)^{N/(N+1)}.
\end{equation}

From Fig.~\ref{fig:emf}, we find that the characteristic momentum scales for $\varepsilon$ and 
$\mu$ are different.
The momentum scale $\kappa$ is related to the temperature scale 
by $T \simeq v(\kappa) \kappa$; therefore, the permittivity $\varepsilon(\kappa)$ 
logarithmically increases below $T_1=v(\kappa_1)\kappa_1$ and 
the permeability decreases below $T_2=v(\kappa_2)\kappa_2$. 
This contrasting behavior helps us to experimentally determine the two characteristic scales. 
In the zero temperature limit, the permeability $\mu = 1 + 4\pi\chi$ 
($\chi$: magnetic susceptibility) goes to zero; i.e., 
the system shows the perfect diamagnetism with $\chi = 1/(4\pi)$.

\subsection{Spectral function}  

The spectral function is obtained as the imaginary part of the electron Green's function, 
so we should carefully select the gauge. To calculate the spectral function, 
we adopt the ``physical gauge,'' i.e., Coulomb gauge.  
The photon propagator in the Coulomb gauge $D^{\mu\nu}_\text{C}(k)$ 
is given by \cite{adkins1983olr} 
\begin{equation}
D^{\mu\nu}_\text{C}(k)=\frac{c^2}{\varepsilon}
\begin{pmatrix}
\dfrac{1}{\bm{k}^2} & 0 \\
0 & -\dfrac{g^{\alpha\beta}}{k^2}-\dfrac{1}{k^2}\dfrac{k^\alpha k^\beta}{\bm{k}^2}
\end{pmatrix}.
\end{equation}

From the Callan-Symanzik equation, the electron Green's function 
$G (\bm{k},\omega)$ is obtained 
as the product of the bare electron propagator, the electron field renormalization $\gamma_2$, 
and the perturbative correction $\mathcal{G}$: 
\begin{equation}
G (\bm{k},\omega)
	=\frac{\mathcal{G}(\alpha(\kappa))}{\omega^2-v^2(\kappa)\bm{k}^2}
	\exp\left[ 2\int_\Lambda^k d\ln\left( \frac{k'}{\Lambda} \right) 
	\gamma_2(\alpha) \right].
\end{equation}
$k$ in this equation should be regarded as a magnitude of a spacelike vector, i.e., 
$k=\sqrt{v^2 \bm{k}^2 - \omega^2}$.  
The field renormalization $\gamma_2$ in the Coulomb gauge is given by \cite{isobe2012tqca}
\begin{equation}
\gamma_2 (v,c,\alpha;\kappa)
	= \frac{\alpha}{\pi}\frac{v^3}{c(c+v)^2}.
\end{equation}

In region (i), the field renormalization $\gamma_2$ is so small with the factor $(v/c)^3$ 
that the correction of the Green's function is negligible. 
On the other hand, the $\kappa$ dependence of $\gamma_2$ in region (ii) is too complicated 
to calculate the Green's function. 
Hence, we concentrate on the analysis for region (iii), where simple analytic expressions exist. 

From Eq.~\eqref{eq:alpha3}, $\gamma_2(k)$ is obtained as
\begin{equation}
\gamma_2 (k) = \frac{\alpha(k)}{4\pi} 
	= \frac{3}{8N}\left[ \ln\left( \frac{\Lambda}{k} \right) \right]^{-1}.
\end{equation}
The perturbative correction $\mathcal{G}$ in region (iii) is small since the running coupling constant 
becomes small in this region, \cite{isobe2012tqca} so we put $\mathcal{G} = 1$.
Then we obtain the electron Green's function
\begin{equation}
G(\bm{k},\omega)=\frac{1}{\omega^2-c_\infty^2\bm{k}^2}
	\left[ \frac{1}{2} 
		\ln \left( \frac{\Lambda^2}{c_\infty^2\bm{k}^2-\omega^2} \right) 
	\right]^{-3/(4N)}.
\end{equation}

The electron spectral function is obtained by the imaginary part of the Green's function 
$-\mathrm{Im} G(\bm{k}, \omega +i0)$. 
It has finite value for $|\omega|\geq c_\infty|\bm{k}|$; otherwise, $-\mathrm{Im}G(\bm{k},\omega)=0$. 
The spectral function in region (iii) has the approximate form
\begin{align}
& -\mathrm{Im}G(\bm{k},\omega) \notag \\
\sim \ &  a\delta(\omega^2-c_\infty^2\bm{k}^2) \notag \\ 
	& + \frac{1}{\omega^2-c_\infty^2\bm{k}^2}
	\frac{3\pi}{8N} 
	\left[ \frac{1}{2} 
		\ln \left| \frac{\Lambda^2}{c_\infty^2\bm{k}^2-\omega^2} \right| 
	\right]^{-1-3/(4N)} \notag \\
	& \times \theta(\omega^2-c_\infty^2\bm{k}^2), 
\end{align}
where the constant $a$ is determined by the sum rule.
The $\delta$ function peak with finite $a$ indicates a Fermi liquid state, which is 
different from the (2+1)D analysis. \cite{gonzalez1994nfl} 
The continuum state for $|\omega| > c_{\infty} k$ is emerged from the electron-electron interaction.

\subsection{Electric conductivity}
\label{sec:conductivity}

In this section, we calculate the electric conductivity for $\omega \ll T$ 
from the quantum Boltzmann equation (QBE) with the leading log approximation.  
Calculations are performed by following previous studies.
\cite{goswami2011qcba,arnold2000tci,fritz2008qct,hosur2012cti} 
The QBE in the external field $\bm{F}$ is 
\begin{equation}
\left[ \frac{\partial}{\partial t} + \bm{F}\cdot\frac{\partial}{\partial\bm{k}} \right]
f_{\lambda a} (\bm{k},t) = -w[f_{\lambda a} (\bm{k},t)],
\end{equation}
where $f_{\lambda a}(\bm{k},t)$ is a distribution function of particles and holes $(\lambda = \pm)$, 
with $a$ being a node index, and $w[f_{\lambda a} (\bm{k},t)]$ represents the scattering rate
due to the electron-electron interaction. 

We assume that the external electric force $\bm{F}=e\bm{E}$ is weak, and that 
the deviation of the distribution function from the equilibrium 
$f^0_{\lambda a} (k) = (1+e^{\beta \lambda v k})^{-1}$ is small, so that 
we consider the linear response in $\bm{E}$:
\begin{align}
f_{\lambda a} (\bm{k},\omega) = & 2\pi\delta (\omega) f^0_{\lambda a} (k) \notag \\
	&+\lambda \frac{\bm{k}\cdot e\bm{E}(\omega)}{k} 
	f^0_{\lambda a} (k) [1-f^0_{\lambda a}(k)] g(k,\omega).
\end{align}

For $\omega \ll T$, the contribution from the particle-hole pair to the current density 
$\bm{j}(\omega)$ can be neglected, thus
\begin{equation}
\bm{j}(\omega) = ev\sum_{\lambda a} \int_{\bm{k}} \frac{\lambda\bm{k}}{k} 
	f_{\lambda a} (\bm{k},\omega). 
\end{equation}
Therefore, the electric conductivity $\sigma(\omega)$ is given by using the function 
$g(k,\omega)$ as 
\begin{equation}
\sigma(\omega) = \frac{j(\omega)}{E(\omega)}
	= e^2v\sum_{\lambda a}\int_{\bm{k}} \frac{k_x^2}{k^2} 
		f^0_{\lambda a}(k) [1-f^0_{\lambda a}(k)] g(k,\omega).
\end{equation}

We should determine $g(k,\omega)$ to obtain the electric conductivity. 
In equilibrium, the scattering rate $w[f^0] =0$, so when we 
expand the scattering rate in terms of $g(k,\omega)$, the zeroth-order term 
vanishes, and we can write 
\begin{equation}
w[f_{\lambda a}(\bm{k},\omega)] 
	= e\bm{E}(\omega)\cdot\mathcal{C}[\lambda g(k,\omega) \hat{\bm{k}}] 
	+ \mathit{O}(g^2),
\end{equation}
where $\hat{\bm{k}}=\bm{k}/k$, and $\mathcal{C}$ is called the collision operator. 
By using the collision operator, the QBE becomes 
\begin{equation}
\left[ i\omega g_a(k,\omega) +\beta v \right] \lambda \hat{\bm{k}}
	f^0_{\lambda a} (k) [1-f^0_{\lambda a}(k)] 
	= \mathcal{C}[\lambda \hat{\bm{k}} g(k,\omega)].
\end{equation}
To solve this equation, it is convenient to use a variational method. 
The variational functional $\mathcal{Q}[g]$ is given by
\begin{align}
\mathcal{Q}[g] = &
	\sum_a \int_{\bm{k}} 
	\left[ i\omega \frac{g^2(k,\omega)}{2} +\beta v g(k,\omega) \right] 
	f^0_{\lambda a} (k) [1-f^0_{\lambda a}(k)] \notag \\
	& - \frac{1}{2} \sum_a \int_{\bm{k}} \lambda \hat{\bm{k}} g(k,\omega) 
	\mathcal{C}[\lambda \hat{\bm{k}} g(k,\omega)] ,
\end{align}
and the stationary point 
\begin{equation}
\frac{\delta\mathcal{Q}[g]}{\delta g} =0
\end{equation}
gives the solution $g(k,\omega)$. 

When we assume the form $g(k,\omega)=k\xi(\omega)$, according 
to Fritz {\it et al.} \cite{fritz2008qct} , 
we obtain the variational functional $\mathcal{Q}[k\xi(\omega)]$ as 
\begin{align}
\mathcal{Q}[k\xi(\omega)] 
= \ &\frac{1}{4\pi^2}\frac{1}{(\beta v)^5} \left\{
	i\omega \frac{7\pi^4}{30}[\xi(\omega)]^2 + 9(\beta v)^2\zeta(3)\xi(\omega) 
	\right\} \notag \\
	&-\frac{1}{2} \frac{2\pi N}{9 \beta^6 v^5} (\alpha^2 \ln\alpha^{-1} )
	F\left( \frac{v}{c} \right) [ \xi(\omega) ]^2,
\end{align}
with the relativistic correction 
\begin{equation}
\label{eq:correction}
F(x) = 1+ \frac{1}{4}\left[ 3-x^2-\frac{(1-x^2)(3+x^2)}{x} \tanh^{-1}x \right].
\end{equation}
The function $F(v/c)$ can be regarded as the relativistic correction, and 
it cannot be obtained from the previous nonrelativistic analyses. 
In the nonrelativistic limit ($v/c \to 0$), we have $F(v/c)=1$, and 
it monotonically increases to $F(v/c)=3/2$ ($v/c \to 1$). 

\begin{figure}
\centering
\includegraphics[width=0.9\hsize]{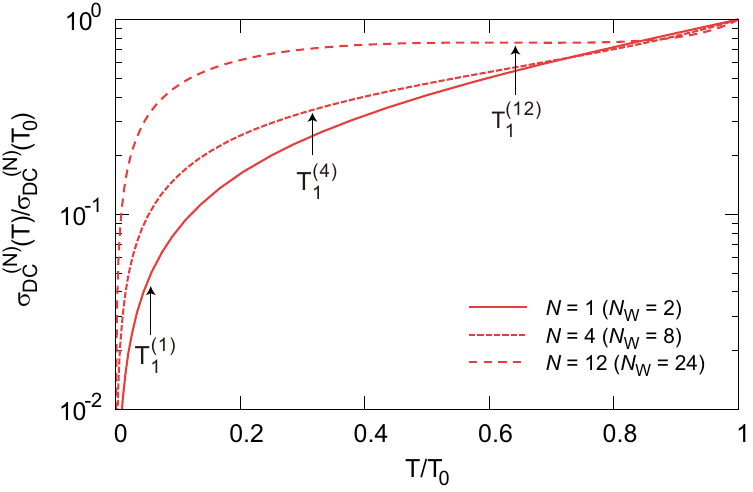}
\caption{%
(Color online) 
DC conductivity $\sigma_\text{DC}^{(N)}(T)$. 
}
\label{fig:RG_cond}
\end{figure}

Now we can determine $\xi(\omega)$ by the functional derivative as
\begin{equation}
\xi(\omega) = \frac{81\zeta(3)}{4\pi^3} \beta^3 v^2
\left[ 2N (\alpha^2 \ln\alpha^{-1}) F\left(\frac{v}{c}\right)
	-\frac{21\pi}{20}i\beta\omega \right]^{-1},
\end{equation}
and the electric conductivity is
\begin{align}
\sigma^{(N)}(\omega,T) = &
2N \frac{e^2}{h^2} \left( \frac{k_\text{B}T}{\hbar v} \right)
	\frac{243[\zeta(3)]^2}{4\pi^4} \notag \\
	& \times \left[ 2N (\alpha^2 \ln\alpha^{-1}) F\left(\frac{v}{c}\right)
	-\frac{21\pi}{20}i\beta\omega \right]^{-1}.
\end{align}
We recovered $k_\text{B}$ and $\hbar$ in the last line of the equation. 
Especially, the DC conductivity is 
\begin{equation}
\sigma_\text{DC}^{(N)}(T) = \frac{e^2}{h}\frac{k_\text{B}T}{\hbar v} 
	\frac{0.90}%
	{\alpha^2 \ln\alpha^{-1} F(v/c)},
\end{equation}
as shown in Fig.~\ref{fig:RG_cond}. 

\subsection{Energy gap}

Finally, let us consider the RG effect on the mass parameter $m$. 
The mass $m$ describes the critical behavior of the gap, and $m=0$ at the 
critical point. 
For particular materials with Dirac nodes, the bare mass $m_0$ is tunable, 
depending on the concentration $x$ or by pressure $P$. \cite{xu2011tpt,sato2011uma} 

The RG equation for mass $m(\kappa)$ is obtained from the electron self-energy as 
\begin{equation}
\kappa\frac{dm}{d\kappa}=-\frac{3\alpha}{2\pi}m, 
\end{equation}
and its analytic solution for $\kappa \gtrsim \kappa_2$ is 
\begin{equation}
\label{eq:mass}
m(\kappa)=m(\Lambda) 
	\left[ 1+\frac{2N+2}{3\pi} \alpha_0
		\ln \left( \frac{\Lambda}{\kappa} \right) \right]^{9/(4N+4)}.
\end{equation}
When we neglect the weak singularity of $\ln\ln m_0$, 
the solution to Eq.~\eqref{eq:mass} becomes
\begin{equation}
m=
m_0 \left[ 1+\frac{2N+2}{3\pi}\alpha_0 \ln \left( \frac{\Lambda}{m_0} \right) 
	\right]^{9/(4N+4)}.
\end{equation}

\section{Discussions and Summary}

Now we discuss the relevance of the present results to the real systems. 

First, for a topological insulator ($N=1$), 
the velocity $v_0$ is estimated at $v_0 \simeq 10^6\,\mathrm{m/s}$ 
from the ARPES measurement of the energy dispersion \cite{xu2011tpt}; 
hence, $c_{\text{vacuum}}/v_0 \simeq 300$.
As for the dielectric constant $\varepsilon$, we take the typical value
$\varepsilon_0 \simeq 10^2$ of BiSb alloys. \cite{qi2009imm}
Since $c_0 = c_{\text{vacuum}}/\sqrt{\varepsilon_0}$, 
$c_0/v_0 \simeq 30$ and $\alpha_0 =  (1/137)/(\varepsilon v) \simeq 0.022$ are obtained.
These values give the estimates for $\kappa_1 \simeq 10^{-47} \kappa_0$
and $\kappa_2 \ll \kappa_1$. 

For the pyrochore iridate Y$_2$Ir$_2$O$_7$ with $2N=24$, 
the velocity and the dielectric constant may be estimated as 
$v_0 \simeq 10^6\,\mathrm{m/s}$ and $\varepsilon_0 \simeq 10$. \cite{hosur2012cti} 
In this case $c_0/v_0 \simeq 95$ and $\alpha_0 \simeq 0.22$, so 
we obtain $\kappa_1 \simeq 0.2 \kappa_0$, and $\kappa_2$ is extremely small. 
The value $\kappa_1 \simeq 0.2 \kappa_0$ would be physically accessible. 

To experimentally observe the RG effects, we have to search materials 
with reasonably large $\kappa_1$ and $\kappa_2$. 
A larger coupling constant $\alpha_0$ is necessary to obtain larger $\kappa_1$, and 
this can be realized if both of the dielectric constant $\varepsilon_0$ and 
the Fermi velocity $v$ are small. 
In addition to large $\alpha_0$, small $c_0/v_0$ is required to make $\kappa_2$ larger. 
There seem to be two approaches: (a) small $c_0$ and (b) large $v_0$. 
In case (a), a large dielectric constant $\varepsilon_0$ leads to the small 
coupling constant $\alpha_0$ (assuming $\mu_0=1$), so it cannot be a solution. 
In case (b), a large $v_0$ also brings a small $\alpha_0$. 
The only way out is the small ratio of $c_0/v_0$. 
Unfortunately, it would be difficult
to observe the relativistic scaling behavior 
at the experimentally accessible temperature in the materials at hand. 

This estimation gives a justification for the nonrelativistic approximation.  
Physically accessible $\kappa_1$ is easily obtained by choosing appropriate 
$v_0$ and $\varepsilon_0$, but it would be difficult to access $\kappa_2$ 
unless $c_0 \approx v_0$. 
It means that the nonrelativistic approximation in the RG analysis is adequate 
in ordinary situations. 
However, if $c_0 \approx v_0$ is accomplished with $\varepsilon_0 \sim 1$ and 
$\mu_0 \gg 1$, we might reach $\kappa_2$, i.e., the relativistic scaling region. 

In summary, we have studied the electromagnetic interaction in (3+1)D 
multi-node ($N$) Dirac systems by using RG analysis. 
The RG equations for the speed of light $c$, that of electron $v$, and the 
coupling constant $\alpha$ are derived for generic $N$.
We solved the RG equations to obtain the analytic expressions for the large $N$ 
limit and the reasonably accurate analytic solutions for generic $N$ systems. 
We also discussed the physical quantities based on the RG analysis, which 
facilitates the observation of the scale-dependent behavior.

\begin{acknowledgments}
We acknowledge fruitful discussions with S. \mbox{Nakosai}. 
This work is supported by Grant-in-Aid for Scientific Research
(Grant No.~24224009)
from the Ministry of Education, Culture,
Sports, Science and Technology of Japan, Strategic
International Cooperative Program (Joint Research Type)
from Japan Science and Technology Agency, and Funding
Program for World-Leading Innovative RD on Science and
Technology (FIRST Program).
\end{acknowledgments}

\bibliography{jabref}

\end{document}